\documentclass{article}
\usepackage{epsf,graphicx}

\newlength{\refindent}
\newlength{\parskiplen}
\begin{document}


\newenvironment{references}{\clearpage
			    \section*{\large \bf REFERENCES}
			    \parindent=0mm \everypar{\hangindent=3pc
			    \hangafter=1}}{\parindent=\refindent \clearpage}
\newenvironment{figcaps}{\clearpage
			 \section*{\large  \bf FIGURE CAPTIONS}}{}
\newcommand{\fig}[2]{\parbox[t]{2.0cm}{Figure #1:} \
		   \parbox[t]{13.5cm}{#2}\\[\baselinestretch\parskiplen]}


\begin{titlepage}
\begin{center}
\vspace*{0.5cm}
{\huge The Detailed Forms of the LMC Cepheid}\\[0.5cm]
{\huge PL and PLC Relations}\\[3.0cm]

{\large C. Koen$^1$, S. Kanbur$^2$ and C. Ngeow$^3$}\\[1cm]
\normalsize
{\em  1 Department of Statistics, University of the Western Cape,
Private Bag X17, Bellville, 7535 Cape, South Africa}\\[0.6cm]
{\em  2 Department of Physics, State University of New York at Oswego, 
Oswego, NY 13126, USA}\\[0.6cm]
{\em  3 Department of Astronomy, University of Illinois, 
Urbana-Champaign, IL 61801, USA}\\[2.0cm]

\end{center}

\begin{quotation}\noindent{\bf ABSTRACT.}
Possible deviations from linearity of the LMC Cepheid PL and PLC relations are
investigated. Two datasets are studied, respectively from the OGLE and MACHO 
projects. A nonparametric test, based on linear regression residuals,
suggests that neither PL relation is linear. If colour dependence is allowed for
then the MACHO PL relation is found to deviate more significantly from the linear,
while the OGLE PL relation is consistent with linearity. These finding are confirmed
by fitting ``Generalised Additive Models" (nonparametric regression functions)
to the two datasets. Colour dependence is shown to be nonlinear in both datasets,
distinctly so in the case of the MACHO Cepheids. It is also shown that there is
interaction between the period and colour functions in the MACHO data. 

\vspace*{1.0cm}
\end{quotation}

\end{titlepage}

\section{INTRODUCTION}

Cepheids are important objects in Astrophysics, both because of their use in the 
extra-galactic distance
scale and their role in stellar evolution. Their regularly repeating 
light curves offer an important opportunity
to test theories of stellar evolution against stellar pulsation: mass-luminosity 
(ML) relations mandated from evolutionary calculations
can be used as input to full linear and non-linear hydrodynamic models of 
Cepheids and compared to observations. These ML relations contain
input about evolutionary physics such as the amount of convective overshoot. 
Constraining theoretical models with observations can be used to gain 
considerable insight into
evolutionary/pulsation physics. On the other hand the Cepheid period-luminosity 
(PL) relation has played an important role in establishing the
extra-galactic distance scale and the subsequent estimation of Hubble's constant, 
$H_0$. The $HST$ Key Project (Freedman et al. 2001) has used $HST$ observations of
Cepheids in a number of galaxies to estimate $H_0$ to within $10\%$ accuracy. 
The crucial step in this work has been the Cepheid PL relation in the
Large Magellanic Cloud (LMC) which has been used to characterize a Cepheid PL 
relation template. This PL template has
traditionally been thought to be linear, however there has also been recent work 
implying a variation of the slope with period in the LMC (Tammann \& Reindl 2002; 
Kanbur \& Ngeow 2004, 2006; Sandage et al. 2004; Kanbur et al. 2007a; 
Ngeow et al. 2005; Ngeow \& Kanbur 2006a,b). 

Ngeow and Kanbur (2006c) estimate the error in estimating $H_0$, if a linear 
Cepheid PL relation is assumed
and the underlying relation is "non-linear"
at a period of 10 days, and find this can lead to an error of about $1-2\%$. 
Such an error seems small but with significant
work being carried out to reduce zero point errors (Macri et al 2006), it is 
important to construct as accurate a distance scale as possible that is independent of
the CMB. Further, table 2 of Spergel et al (2007) points to the fact that an 
independent estimate of $H_0$, accurate to less than
$5\%$, will help to break the degeneracy between ${\Omega}_{matter}$ and $H_0$ 
present from WMAP CMB studies. An independent estimate
of $H_0$ accurate to $1\%$ will result in a reduction of the $65\%$ confidence 
interval on ${\Omega}_{matter}$ by almost a factor of two over
that with WMAP data alone. 

In previous studies, a rigorous statistical test, the $F$ test, was 
applied to the LMC Cepheids to test for the linear versus non-linear 
PL relation. Here by ``non-linear'' we mean two lines of significantly 
differing slope which are continuous at a period of 10 days. The $F$ test
 results that were obtained from the OGLE (Optical Gravitational Lensing Experiment, 
Udalski et al. 1999) and MACHO Cepheid data, in Kanbur \& Ngeow (2004; 2006) 
and Ngeow et al. (2005) respectively, strongly imply that the LMC PC/PL 
relations are non-linear. It is important to note that several other 
statistical tests, such as the ${\chi}^2$ tests, least absolute deviation, 
robust estimation and loess procedures, were also applied to the MACHO data, 
and these results also point to a non-linear LMC PL relation 
(Ngeow et al. 2005). Recently, Kanbur et al (2007a) developed the use of 
testimators and a likelihood based method using the Schwarz Information 
Criterion, to study non-linearities in the LMC PL relation (using both 
OGLE and MACHO Cepheid data) and again came to the same conclusion: the 
LMC Cepheid PL relation is non-linear in the sense described above. The 
$F$ test also suggested that the LMC period-colour (PC) relation is 
non-linear, in contrast to the Galactic and SMC (Small Magellanic Cloud) 
PC relations (Kanbur \& Ngeow 2004). Since the question of the non-linearity 
of the LMC PL relation is important in distance scale and stellar studies, 
it is vital to establish this as firmly as possible; this is one of 
the motivations for this paper.

In addition to investigating the non-linearity of the LMC PL relation, we 
also study the LMC period-luminosity-colour (PLC) relation.
A number of authors, including Sandage (1958) and Madore and Freedman (1991) 
have derived the Period-Luminosity-Color (PLC) relation and shown how it arises from 
the period-mean density theorem, the Stefan-Boltzmann law and the existence of 
an instability strip. These authors also point out that the PL/PC relations are 
obtained from the PLC relation by averaging over the variable not included
in the relation. 

In Section 2, we briefly describe the 
data used in our study. In Section 3 we apply a preliminary test study 
on the LMC PL relation. This is followed by more detailed analysis in Section 4,
based on a non-parametric model fitting procedure. 
An extension to the PLC relation is presented in Section 5. 
The conclusion and discussion of our results are given in Section 6.

We add a few sentences on the use of non-parametric 
methods in what follows. The term ``non-parametric" is actually used in three slightly 
different senses. First, the major innovation (sections 4 and 5) in this paper is the 
use of ``non-parametric regression". The meaning is {\it not} necessarily the usual 
one of ``distribution-free": rather, it means that the form of the regression is not
specified -- the regression function is ``unstructured", being dictated by the data
itself. Of course, this flexibility allows one to detect subtleties which may 
otherwise be overlooked. Second, in the next section of the paper we use a well-known 
distribution-free statistic, the ``Wald-Wolfowitz runs test". This non-parametric 
statistic uses only data ranks, and hence typically not very powerful. Third, also 
in the next section use is made of a permutation method. This avoids 
distributional assumptions about the data, by 
using re-orderings of the data itself to establish significance levels.  

\section{THE DATA}

We use two sets of LMC Cepheid data in our study. The first data set is the 
extinction corrected $V$-band mean magnitudes and $(V-I)$ colours for the OGLE 
LMC Cepheids taken from Kanbur \& Ngeow (2006), supplemented with additional 
Cepheids from Sebo et al (2002), and referred as ``OGLE'' data in this paper. 
The second data set is the MACHO Cepheids data, with extinction corrected $V$ 
mean magnitudes and $(V-R)$ colours, adopted from Ngeow et al (2005).
Using these two data sets allow us to compare the results, particularly for the 
different photometric filters used.

A possible complication is that any apparent non-linearity in PL or PLC
relations could be caused by 
extinction errors which are a function of colour or period.
Arguments against extinction errors as a cause of observed non-linear LMC PL 
and PC relations were presented in Kanbur \& Ngeow (2004), Kanbur \& Ngeow (2006),
Kanbur et al. (2007b), Ngeow et al. (2005), Ngeow \& Kanbur (2006b) and
Sandage et al. (2004), and will therfore not be repeated in detail here.
In particular, a possible period dependency of extinction errors has been 
investigated in Ngeow \& Kanbur (2006b). If such extinction errors were present, 
then the PC relations at maximum light would be such that LMC Cepheids would get 
hotter at maximum light as the pulsation period increases: a fact which would
be hard to reconcile with pulsation theory especially as Galactic Cepheids, 
in common with LMC Cepheids, display a flat PC relation at
maximum light (Kanbur \& Ngeow 2004, 2006). Further, the dependence of extinction 
error on colour would need to
be very complicated to explain both the non-linearity at mean light whilst 
preserving the flatness at maximum light. 

It is also noted that the reddening values adopted here are the {\it same} 
as those used in many distance scale studies (Freedman et al. 2001).

\section{A PRELIMINARY INVESTIGATION BASED ON A TEST PROCEDURE}

Figs. 1 and 2 show the MACHO and OGLE PL data, with least squares
linear fits of the form
\begin{equation}
V=a+b \log P +{\rm error} \; .
\end{equation}
For the sake of completeness,
\begin{eqnarray}
V&=&17.08(0.026)-2.70(0.039) \log P \;\;\;\;\;(MACHO)\nonumber\\
V&=&17.05(0.020)-2.69(0.028) \log P \;\;\;\;\;\;(OGLE)
\end{eqnarray}
where standard errors of coefficient estimates are given in brackets.
Although both fits are excellent, it is nonetheless
of some interest whether there may be subtle deviations from the strictly
linear relations between $V$ and $\log P$ shown by the lines:
although this may have little importance for prediction of luminosity
given the period, it could (e.g.) have an important bearing on the modelling
of Cepheid pulsations.

A simple procedure which provides some insight into the problem is
to study partial sums of the residuals of the least squares fits. 
First arrange the data so that the period values are in ascending order:
$$P_1 <P_2<P_3<\ldots <P_N $$
where $N$ is the sample size. Then
\begin{equation}
C(j)=\sum_{k=1}^j [V_k-a-b \log P_k]=\sum_{k=1}^j r_k
\end{equation}
are the partial sums of the residuals $r_k$. If there are no deviations 
from linearity, then $C(j)$ is the sum of uncorrelated random numbers and 
hence a simple random walk. However, if there are deviations 
from linearity successive residuals may be correlated, and hence $C(j)$
will not be a simple random walk. Partials sums of the $r_k$ can be
seen in Figs. 3 and 4.

A statistic which can be used for testing whether the partial sum
is a pure random walk is its vertical range
$$R=\max_j C(j)-\min_j C(j) \; :$$
this may be expected to be inflated by positively correlated residuals. Significance
levels for the values of $R$ are readily obtained by permutation, as 
follows:
\begin{itemize}
\item[(i)]
Permute the $r_k$; this will randomise the residuals by destroying any
possible trends.
\item[(ii)]
The partial sums of the permuted $r_k$ will be true random walks -- 
find the statistic $R$ for the permutation.
\item[(iii)]
Repeat steps (i) and (ii) a large number of times, noting the values of
$R$. 
\item[(iv)]
Determine the fraction of permutation $R$-values which exceeds the observed
value -- this estimates the significance level of the observed $R$.
\end{itemize}
Applying 10000 permutations, significance levels of 3\% and 4\% were
obtained for the MACHO and OGLE data respectively, suggesting 
meaningful deviation of the observed $r_k$ from randomness. The implication
is therefore that the PL relation is not perfectly linear.

Study of Figs. 3 and 4 shows that there is an excess of positive residuals 
for $\log P \sim 0.5$ and $\log P>1$, and an excess of negative values
for $0.8<\log P<1$.

Interestingly, application of the standard Wald-Wolfowitz runs test
(e.g. Conover 1971) for randomness of the residuals gives conflicting results
for the two datasets -- significance levels of 45\% and 0.9\% for
the OGLE and MACHO data respectively. Of course, the procedure
uses only the signs, and not the sizes, of the $r_k$.

It is known that Cepheids follow a PLC, rather than simply a PL,
relation. It may therefore be prudent to replace (1) by 
\begin{equation}
C(j)=\sum_{k=1}^j [V_k-a-b \log P_k-c(CI)_k]
\end{equation}
where $(CI)$ indicates a colour index, with regression coefficient
$c$. This has a substantial influence on the significance levels 
of the statistic $R$: for the OGLE data is increases to 33\%, while
the level for the MACHO data is reduced to 0.7\%. The corresponding
Wald-Wolfowitz test levels are 43\% and 1.5\%. 

To summarise, there is strong evidence of non-randomness in the residuals
of the MACHO data, both for the PL and the PLC relations. For the OGLE
data the results are ambiguous.

\section{PL RELATION}

An alternative to the imposition of a fully specified parametric
model such as (1) is to allow the form of the regression to be
dictated by the data. The idea is conveniently illustrated by
a technique known as ``loess" (see e.g. Cleveland \& Devlin 1988). 
Ngeow et al (2005)
initially used this method on MACHO data and found a similar result to 
that reported here. Here we study it in more detail and apply it to both 
MACHO and OGLE Cepheid data.
The method entails fitting a low order polynomial (in the present
case a straight line) over restricted sections (``windows") of the
data by weighted least squares. In the implementation here the only
free parameter is the width $\alpha$ of the window, which is usually 
given as a fraction of the range of the independent variable (i.e.
$0<\alpha \le 1$) . The smaller $\alpha$
the more ``local" the estimated regression, and the more detail 
it shows. Fig. 5 shows a loess regression of the OGLE data, using
$\alpha=0.05$; if $\alpha$ is increased towards unity the loess regression
resembles the linear fit of Fig. 2. 

A key element is then obviously the choice of window width $\alpha$, and
it is desirable to use an objective method to find it. This is readily
done by ``cross-validation": 
\begin{itemize}
\item[(i)]
Choose a value of the window width $\alpha$.
\item[(ii)]
Leave out the first datapoint and obtain a loess estimate  
$\widehat{V}_1$ of the magnitude $V_1$
by fitting the regression to the remaining data.
\item[(iii)]
Note the discrepancy 
$$\Delta_1=V_1-\widehat{V}_1$$
between the true and predicted values.
\item[(iv)]
Repeat steps (ii)-(iii) for the second, third,..., last datapoints,
giving the set $\Delta_1, \Delta_2,\ldots,\Delta_N$ of discrepancies.
\item[(v)]
The value of the cross-validation criterion for the value of $\alpha$
from (i) is the defined as
\begin{equation}
CV(\alpha)=\frac{1}{N} \sum_{j=1}^N \Delta_j^2
=\frac{1}{N} \sum_{j=1}^N (V_j-\widehat{V}_j)^2
\end{equation}
Clearly, it evaluates the predictive power over all the observations
of the loess fit based on the particular value of $\alpha$.
\item[(vi)]
Repeat steps (i)-(v) for all candidate values of $\alpha$.
\item[(vii)]
The optimal $\alpha$ is that which minimises $CV(\alpha)$.
\end{itemize}

The cross-validation functions for the two datasets are plotted in Fig. 6;
optimal window widths are 0.36 and 0.20 
respectively for the MACHO and OGLE observations. In Figs. 7 and
8 the resultant loess
functions are compared to the regression lines from (1). A small difference
between the curves over the approximate interval $0.8<\log P<1$ is visible
in both diagrams. There is also a substantial disagreement at the longest
periods for the MACHO results in Fig. 7: this is clearly due to the 
{\it systematic} difference between the data and the linear regression
line for $\log P>1.25$ (see Fig. 1). Similarly, the slight divergence 
between the loess and linear regression lines at the longest periods
in Fig. 8, can be traced to the influence of the two OGLE datapoints with
$\log P>1.7$ (see Fig. 2).

The question arises as to whether the discrepancies between the loess
curves and the straight line fits are at all meaningful. In order
to address this issue confidence intervals for the loess curves are 
estimated by bootstrapping (e.g. Efron \& Tibshirani 1993). The results,
based on 5000 bootstrap samples, are plotted in Figs. 9 and 10. 
Rather than showing the linear regression line and the 95
lower limits, the {\it difference} between the linear fit and the
confidence limits are plotted, in order to more clearly display
the deviations. It is notable that the linear fits lie outside the
confidence intervals for the loess functions for $0.8<\log P<1$ roughly.
This supports previous work which has suggested a "break" around a 
period $\log P \approx 1$
(Kanbur \& Ngeow 2004, Ngeow et al 2005, Kanbur et al 2007a).

The {\textsc R} software add-on package ``mcgv" contains an alternative 
nonparametric regression facility in the form of thin plate regression 
splines (TPRS) (e.g. Wood 2006). The form of cross-validation used is based
on a balance between the sum of squared model residuals (which measures
the goodness of the model fit) and a smoothness term. Cross-validation in
mcgv is automated. 

The loess and TPRS results are compared for the MACHO and OGLE respectively
in Figs. 11 and 12. The agreement is very good -- in particular, the deviations 
from linearity for $0.8<\log P<1$ are also evident in the TPRS results.
Despite the fact
that more effective degrees of freedom are required for the nonparametric
fits (6.41 and 8.71 for the TPRS fits to the MACHO and OGLE data respectively)
than for linear regression (3 degrees of freedom), the former fits follow
the data considerably more closely. Model selection tools such as the
``Akaike Information Criterion" (AIC, e.g. Burnham \& Anderson 2002) can be used to test 
whether the improved model fit warrants the additional degrees of freedom expended.
In this case, the TPRS fits are both preferred by very wide margins.

\section{PLC RELATION}

Unusual datapoints can have substantial, often somewhat distorting,
influences on regression surfaces. It is therefore worthwhile examining
the datasets carefully in order to identify such data. This is most
easily done using ordinary multiple linear least squares regression.

Fitting PLC relations to the two datasets give the results 
\begin{eqnarray}
V&=&16.23(0.026)-3.30(0.029) \log P +3.95(0.093)(V-R)\;\;\;\;\;(MACHO)
\nonumber\\
V&=&15.97(0.025)-3.23(0.018) \log P +2.30(0.049)(V-I)\;\;\;\;\;\;(OGLE)
\end{eqnarray}
with residual standard deviations 0.164 and 0.097 mag.
Regression diagnostics were examined in order to identify observations
which gave rise to large residuals and/or were unduly influential on
parameter estimates. ``Cooks's $D$" statistic was used for the latter
purpose -- see e.g. Montgomery, Peck \& Vining (2001) (or almost any
other modern text devoted to linear regression theory). Three points were 
eliminated
from the MACHO data, and four from the OGLE data, on the basis of these 
diagnostics.
The PLC relations were then re-estimated for the reduced datasets, and the
new sets of diagnostics examined. This led to a further two deletions from
the OGLE data. The final results, replacing (6), are 
\begin{eqnarray}
V&=&16.23(0.026)-3.32(0.029) \log P +4.00(0.092)(V-R)\;\;\;\;\;(MACHO)
\nonumber\\
V&=&15.89(0.021)-3.29(0.015) \log P +2.48(0.041)(V-I)\;\;\;\;\;\;(OGLE)
\end{eqnarray}
with residual standard deviations of 0.162 and 0.074 mag. The substantial 
reduction in residual variance, and large changes in regression 
coefficients for the OGLE results are particularly striking.

It is interesting to examine the positions of the rejected observations in
three-dimensional dataplots. The plots in Figs. 11 and 12 were obtained by 
selecting perspectives which clearly show the positions of all questionable
data. It is clear the observations for each dataset lie close to a plane, 
and that points with unsatisfactory regression diagnostics (marked by 
squares) all deviate from the plane. The fact that the plane in Fig. 12 
(OGLE data) is so well-defined explains why removal of the outlying points 
made such a substantial difference to the estimated coefficients. In the 
remainder of this paper we work with the reduced datasets ($N=1213, 717$ 
for MACHO and OGLE data respectively). Note that one high-influence datum
in the OGLE data is retained (for the brightest Cepheid -- see Fig. 12), 
since its associated residual is very small, and since its omission
has very little influence on the values of the three estimated parameters.

An obvious extension of the linear PLC relation to the nonparametric case 
is the so-called ``Generalised Additive Model"
\begin{equation}
V=\alpha+f_P(\log P)+f_C(CI)+{\rm error} 
\end{equation}
where $\alpha$ is a constant; $CI$ denotes a colour index; 
and $f_P$ and $f_C$ are nonparametric 
regression functions such as loess or TPRS fits. Due to the several
attractive features (automated cross-validation, to mention but one)
the {\textsc R} add-on package is once again used to perform
TPRS fits of (8) to the data.

The results can be seen in Figs. 15  and 16. The estimated $f_P$ for the 
OGLE data is linear: the effective degrees of freedom, 1.00, confirms
this. By implication the model (8) reduces to
\begin{equation}
V=\alpha+\beta \log P+f_C(CI)+{\rm error} \;.
\end{equation}
Not surprisingly, the AICs of models (8) and (9) 
are exactly equal for the OGLE data.

The function $f_P$ for the MACHO data shows the familiar deviation from
linearity in the range $0.8<\log P<1$; this is more clearly demonstrated
in Fig. 17, where a linear fit to $f_P$ has been subtracted.

Inspection of the $f_C$ functions in Fig. 16 shows that both are 
distinctly nonlinear.

It is of obvious interest to investigate why $f_P$ reduces to the perfectly
linear form in the case of the OGLE data, when the dependence of $V$ on 
$\log P$ in the PL relation is nonlinear. Examining the relationship between
$\log P$ and the colour index $(V-I)$ gives some insight into this question.
The results of a loess regression of $(V-I)$ on $\log P$ for the OGLE
data are displayed in
Fig. 18. The 95\% confidence intervals, obtained from 5000 bootstrap
samples, are also shown. Calculations were done using a smoothing 
window of width 0.20, as indicated by cross-validation. The analogous 
plot for the MACHO data, based on a smoothing window width of 0.33,
is in Fig. 19. In the case of the OGLE data there is a
clear change in the relationship between $\log P$ and $(V-I)$ 
in the neighbourhood $0.8<\log P<1$. It appears that small deviations from
linearity in the $PL$ relation in Fig. 8 are compensated by the colour dependence.
In the case of the MACHO data the kink in the $PC$ plot (Fig. 19) is of similar 
size to that in Fig. 18, but the deviation from linearity in the $PL$ plot is
larger (Fig. 7). This may explain why the $f_P$ function remains nonlinear
in the case of the MACHO $PLC$ relation. These results support similar work
presented in Kanbur and Ngeow (2004) and Ngeow and Kanbur (2005) 
on the non-linearity of the LMC PC relation using $F$ 
tests, and on the linearity of the LMC Wessenheit function.

Nonparametric regression lends itself to much more flexible forms than
ordinary multiple regression. Two possible alternatives to (8) are
\begin{equation}
V=\alpha+f_P(\log P)+f_C(CI)+f_{PC}(\log P, CI)+{\rm error} 
\end{equation}
and
\begin{equation}
V=\alpha+f_{PC}(\log P,CI)+{\rm error} 
\end{equation}
which allows for interaction between the two independent variables. 

The two Generalized Additive Models (10) and (11) were 
also fitted to both datasets.
For the OGLE data, the AIC-preferred model is (10), but 
a more detailed analysis (ANOVA) shows that the contribution from
the interaction function $f_{PC}$ is not significant -- hence the
model effectively reduces to (8). For the MACHO data the pure
interaction model (11) is preferred, with (10) the second choice. 
According to the AIC, the additive model (8) is a very distant 
third choice. A contour plot of the fit of the model (11) can
be seen in Fig. 20 -- this demonstrates why (8) is inadequate.
Of course, in practice (11) would be more tedious to work with
than the simpler additive form (8).

A few words of explanation of Fig. 20 may be in order. The form of
a purely linear PLC relation would of course be
$$V=a+b\log P+c CI+{\rm error} \; .$$
One way of displaying this graphically would be to draw the
lines 
$$V={\rm constant}$$
in the $\log P$-$CI$ plane, for various values of the constant.
The equations describing these contour lines are 
$$CI=(V-b\log P-{\rm constant})/c +{\rm error} \; ,$$
i.e. straight lines with slope $-b/c$. Fig. 20, the
equivalent for the non-parametric function $f_{PC}$, shows
not only that the relations are nonlinear, but also that there
is ``interaction" -- the form of the relation depends on the region
of the $\log P$--$(V-R)$ plane it inhabits.

\section{CONCLUSIONS \& DISCUSSION}

It should perhaps come as no surprise that with the acquisition of
large amounts of new data finer detail in the relationships between
astrophysical observables are uncovered. 
The best-fitting models of the two datasets are
given by (11) (MACHO) and (9) (OGLE) respectively, which both
are both nonlinear.

Estimates of the effect of such small non-linearities 
on the Cepheid distance scale and
on Hubble's constant are given in Ngeow and Kanbur (2006c) and amount to $1-2\%$. 
Such an error seems small but in the
era of "precision cosmology" with a drive toward a distance scale accurate to 
$5\%$, such an effect is important. Perhaps just
as important, a proper characterization of the precise detail in the observed 
phenomena will assist in placing improved constraints
on pulsation models of Cepheids and in particular on their ML relations, and 
hence on details of stellar evolutionary
physics such as the amount of convective core overshoot.

A possible physical explanation for this non-linearity is outlined in the 
papers by Kanbur et al. (2004), Kanbur \& Ngeow (2006) and
Kanbur et al. (2007b), which studied
Galactic, LMC and SMC Cepheid models respectively. Briefly, these papers 
suggest the non-linearity is caused by the interaction of the hydrogen
ionization front (HIF) and photosphere and the way this interaction varies with 
period. At low densities, if the HIF and photosphere are engaged
(i.e. the photosphere lies at the base of the HIF) then the temperature of 
the photosphere and hence the colour of the star are almost independent
of global stellar properties such as the period. Since the relative location 
of the photosphere and HIF varies with the $L/M$ ratio, and since this
varies with period, modelling has implied that for LMC Cepheids with a period 
greater than 10 days, the photosphere and HIF are not engaged. Thus these
stars have a different PC relation than their shorter period counterparts, 
Because the PC and PL relations are really forms of the PLC relation, then
a change in the PC relation results in a change in the PL relation. 
Galactic Cepheids are such that the HIF-photosphere interaction only really 
occurs at maximum light at low densities. LMC Cepheids are such that this 
HIF-photosphere interaction starts to occur at low densities only
for Cepheids with periods greater than 10 days. SMC Cepheids are such that 
this HIF-photosphere interaction always occurs at high densities (Kanbur et al. 
2004; Kanbur \& Ngeow 2006; Kanbur et al. 2007b). 

\section*{\large \bf ACKNOWLEDGMENTS}
The authors are grateful for the efforts of those who have developed
and maintained the {\textsc R} statistical software. SMK acknowledges 
support from a small research grant from
the American Astronomical Society and the Chretien International research grant. 
CN acknowledges financial support 
from NSF award OPP-0130612 and a University of Illinois seed funding 
award to the Dark Energy Survey.

\pagebreak

\pagebreak

\begin{figure}
\epsfysize=8.0cm
\epsffile{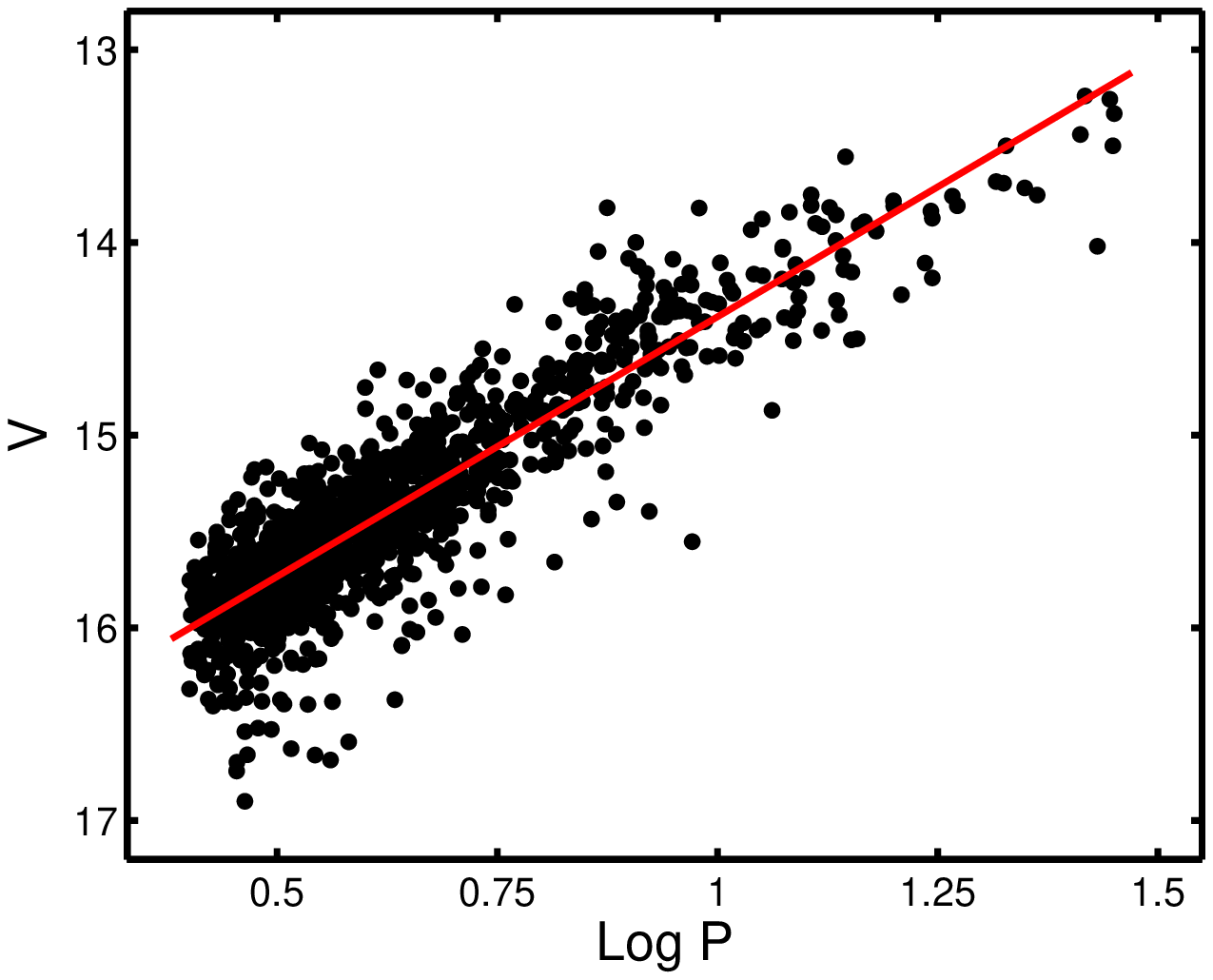}
\caption{MACHO PL data for LMC Cepheids. The line is a linear
least squares fit to the data.}
\end{figure}

\begin{figure}
\epsfysize=8.0cm
\epsffile{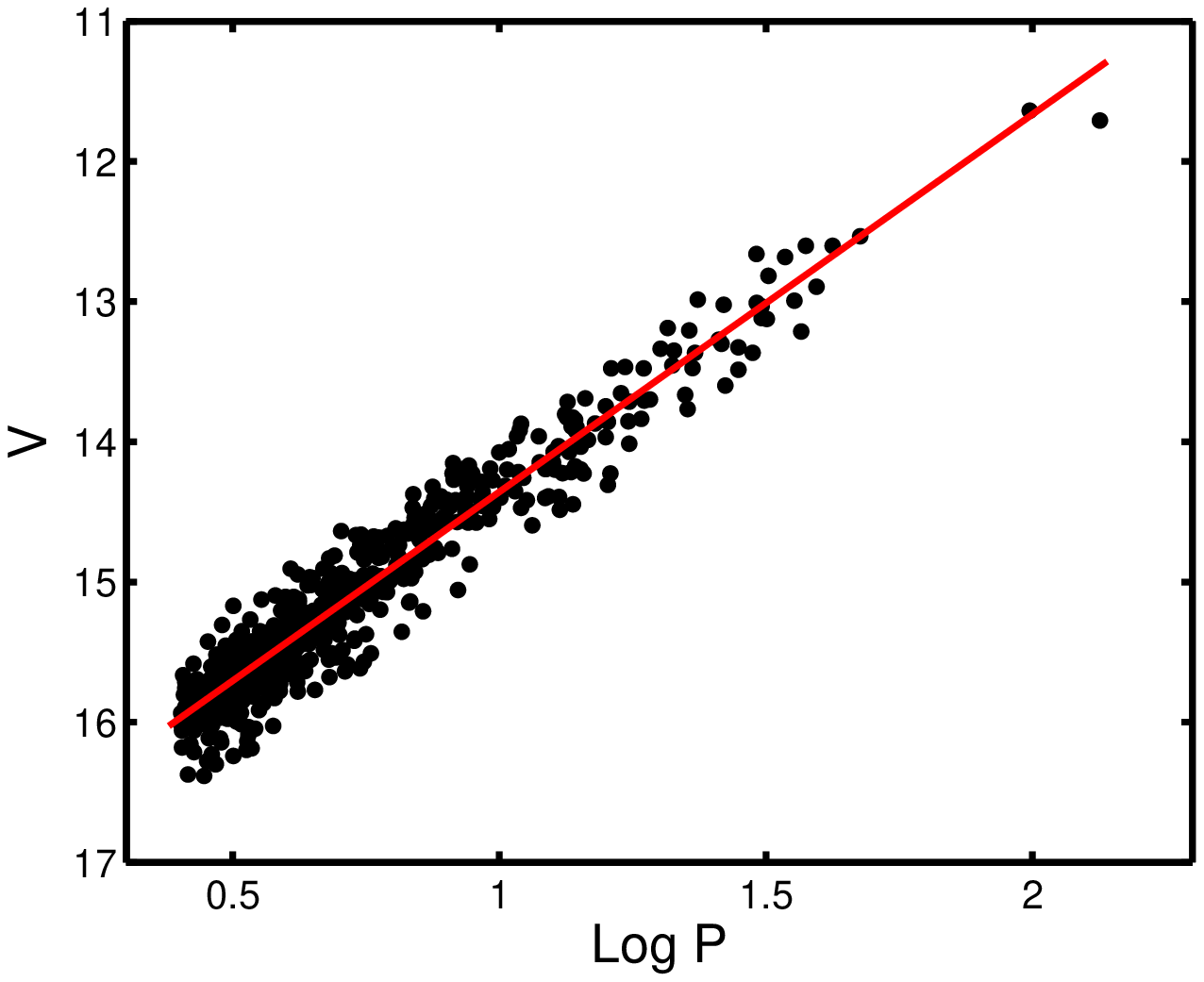}
\caption{OGLE PL data for LMC Cepheids. The line is a linear
least squares fit to the data.}
\end{figure}

\begin{figure}
\epsfysize=8.0cm
\epsffile{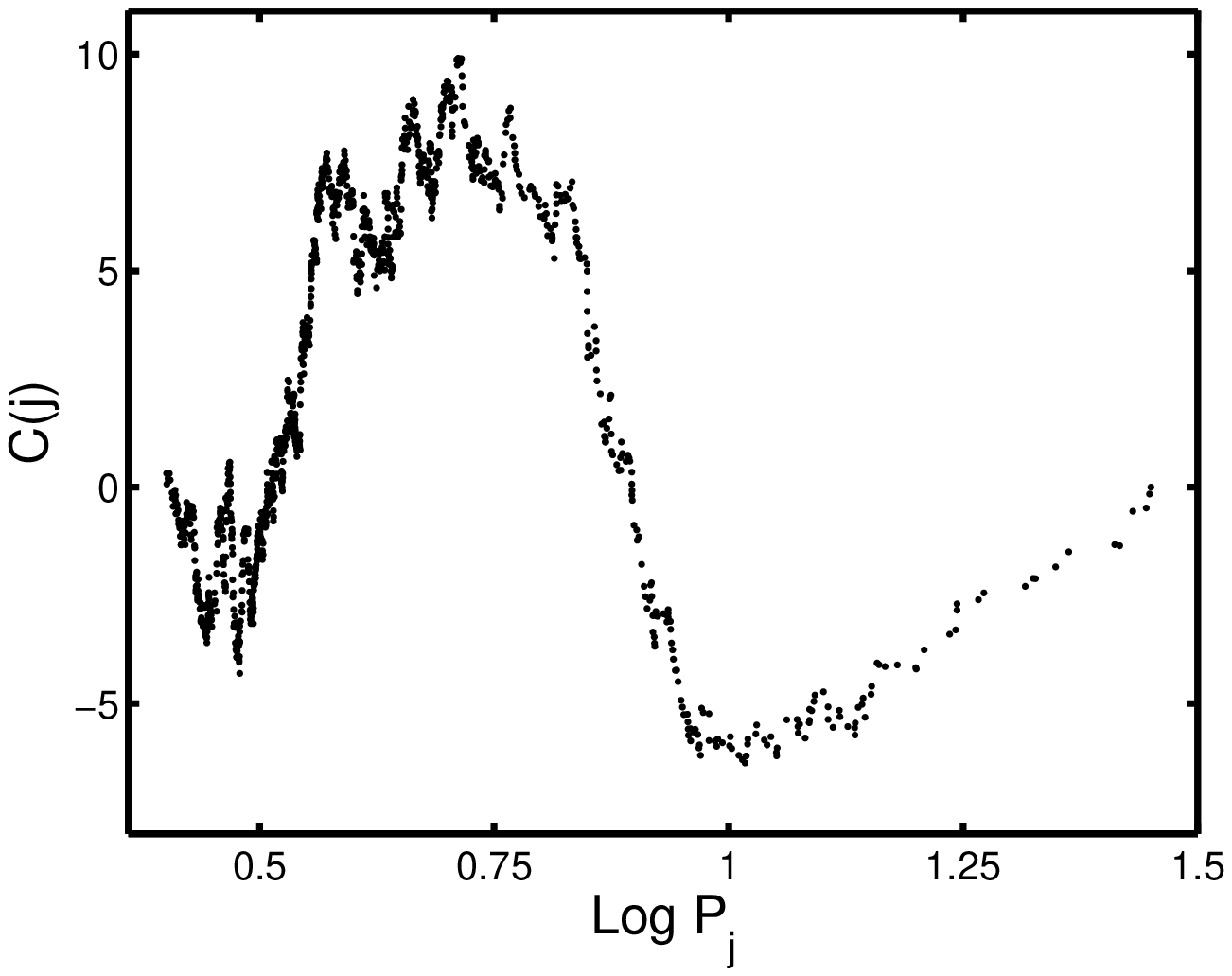}
\caption{Partial sums of the residuals from the fit in Fig. 1.} 
\end{figure}

\begin{figure}
\epsfysize=8.0cm
\epsffile{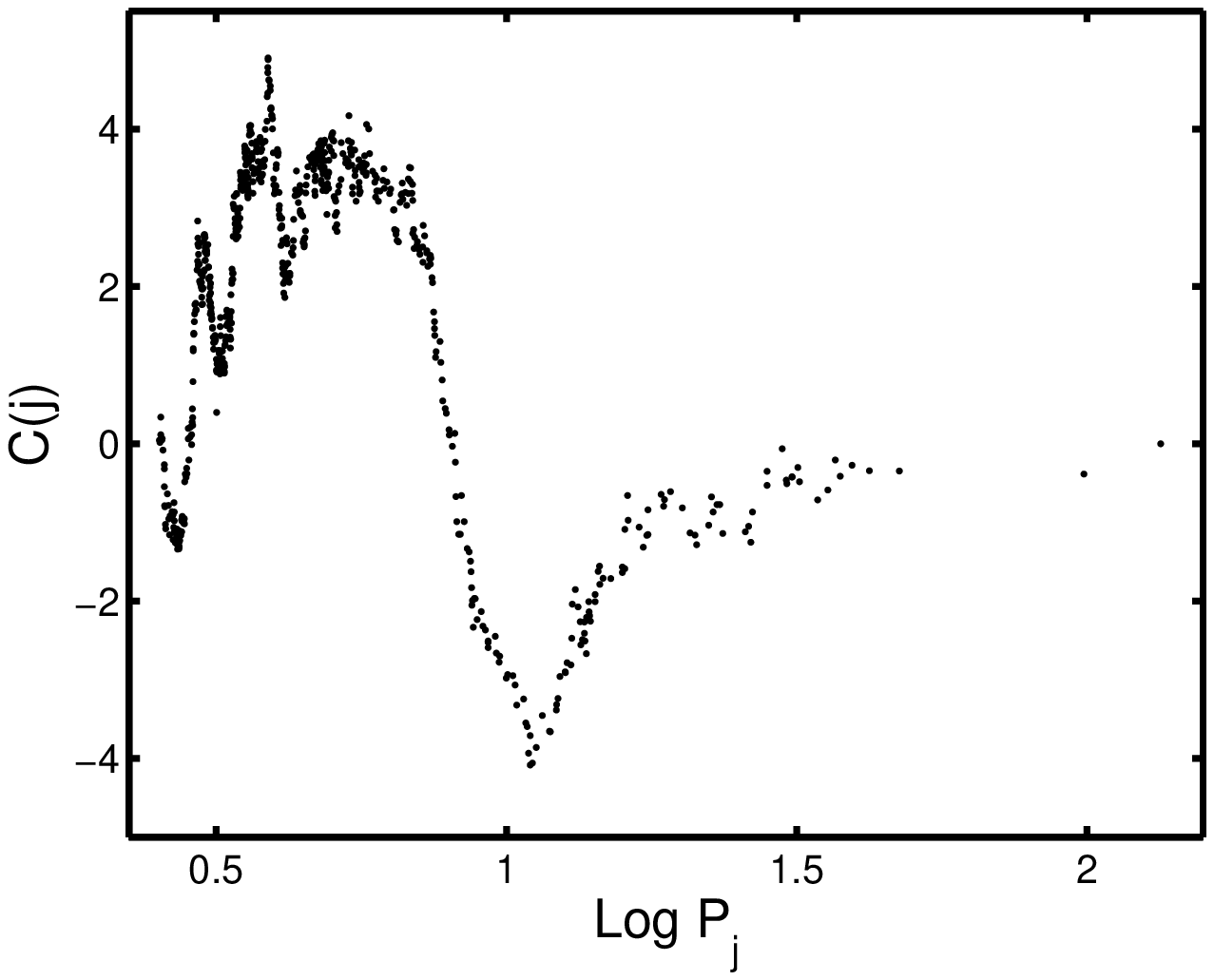}
\caption{Partial sums of the residuals from the fit in Fig. 2.} 
\end{figure}

\begin{figure}
\epsfysize=8.0cm
\epsffile{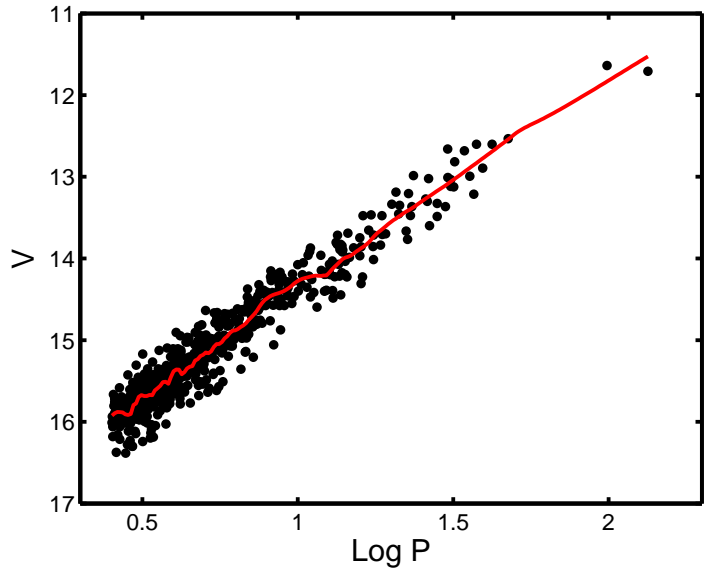}
\caption{An illustrative loess regression on the OGLE PL data. The window
width is 0.05, i.e. 5\% of the range of $\log P$.} 
\end{figure}

\begin{figure}
\epsfysize=9.0cm
\epsffile{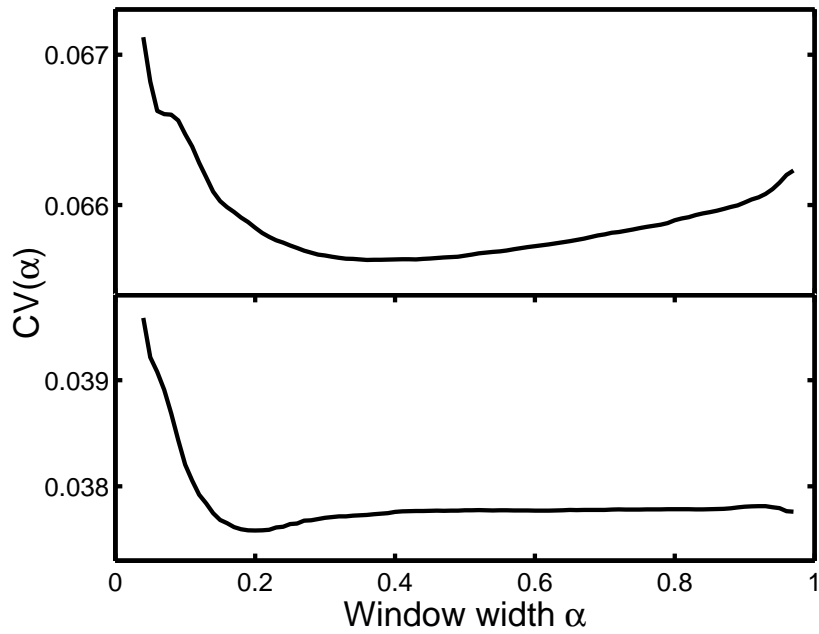}
\caption{Cross-validation functions for the loess window width $\alpha$,
for the MACHO (top) and OGLE (bottom) data.}
\end{figure}

\begin{figure}
\epsfysize=8.0cm
\epsffile{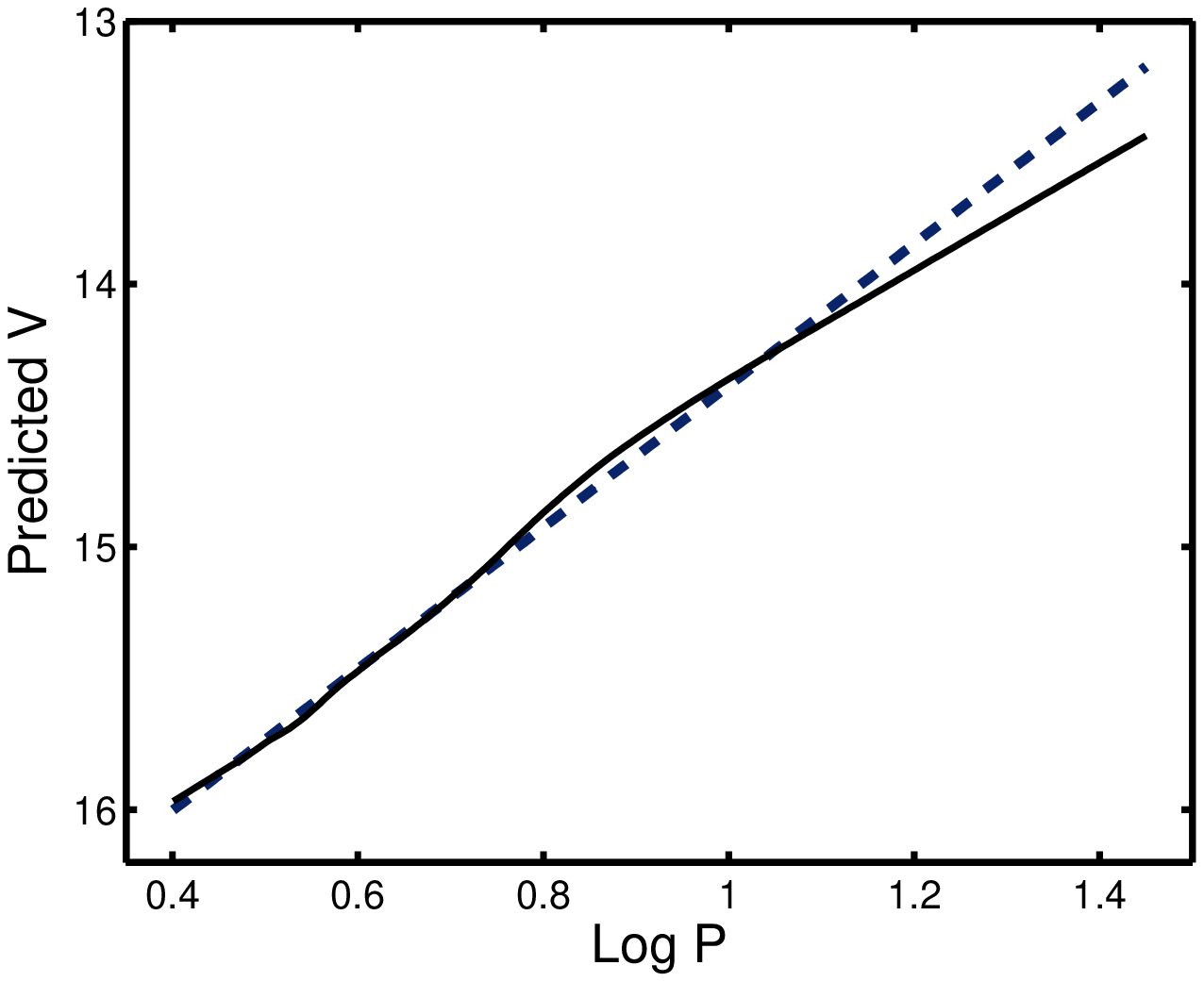}
\caption{A comparison of the optimal loess fit to the MACHO data, and
the linear regression from (1).}
\end{figure}

\begin{figure}
\epsfysize=8.0cm
\epsffile{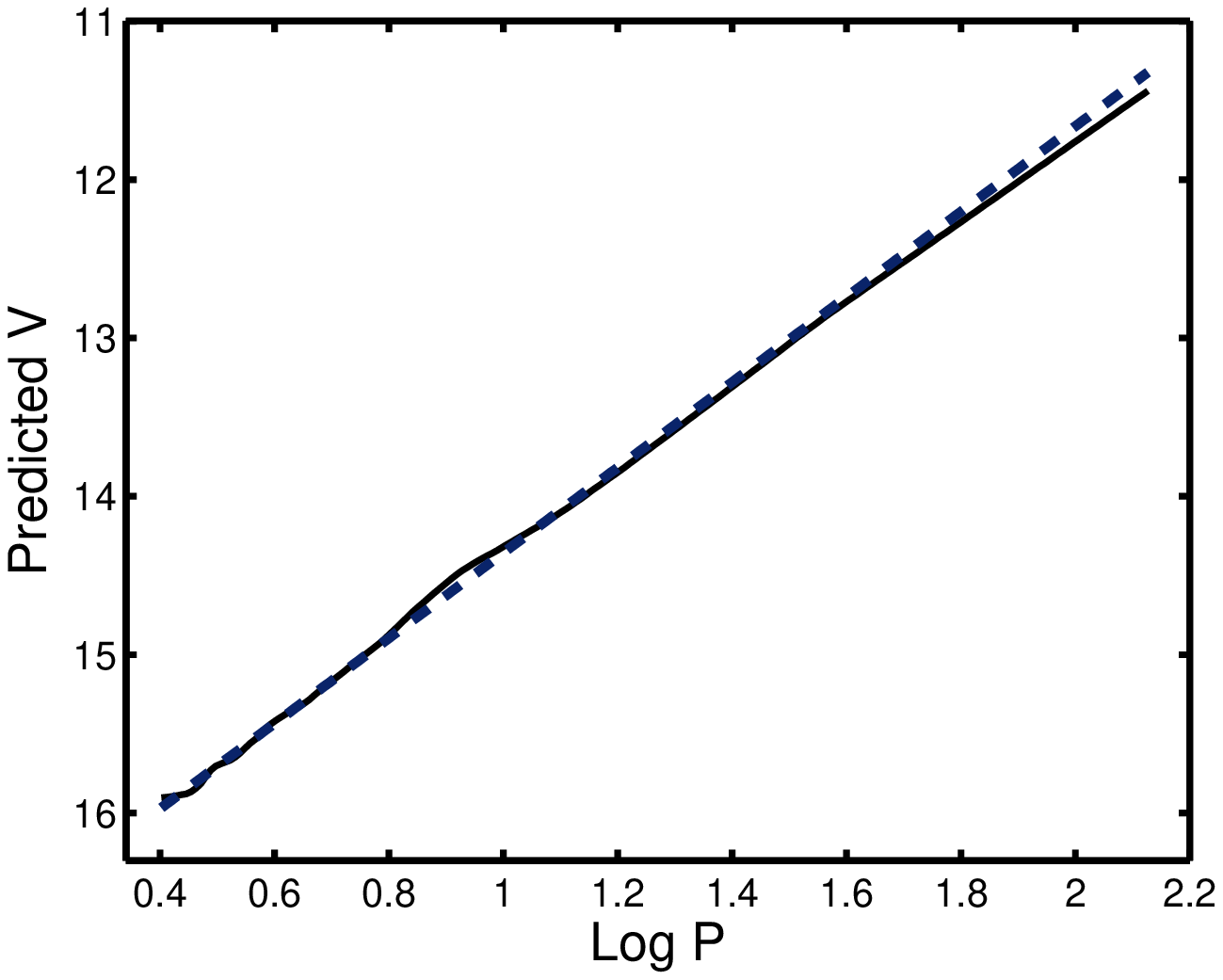}
\caption{A comparison of the optimal loess fit to the OGLE data, and
the linear regression from (1).}
\end{figure}

\begin{figure}
\epsfysize=8.0cm
\epsffile{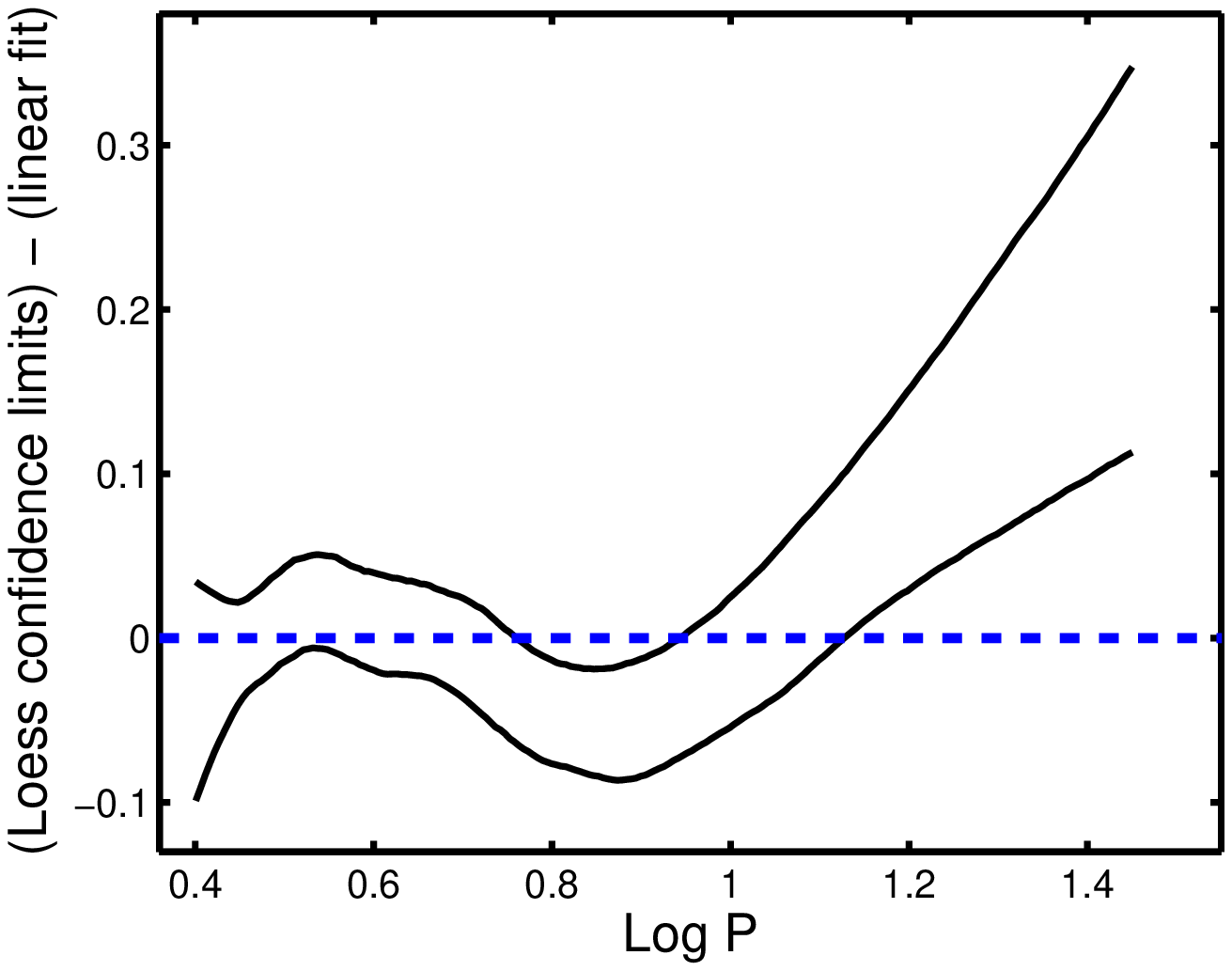}
\caption{The positions (with respect to the linear regression line) 
of the upper and lower 95\% confidence limits
on the loess fit to the MACHO data.}
\end{figure}

\begin{figure}
\epsfysize=8.0cm
\epsffile{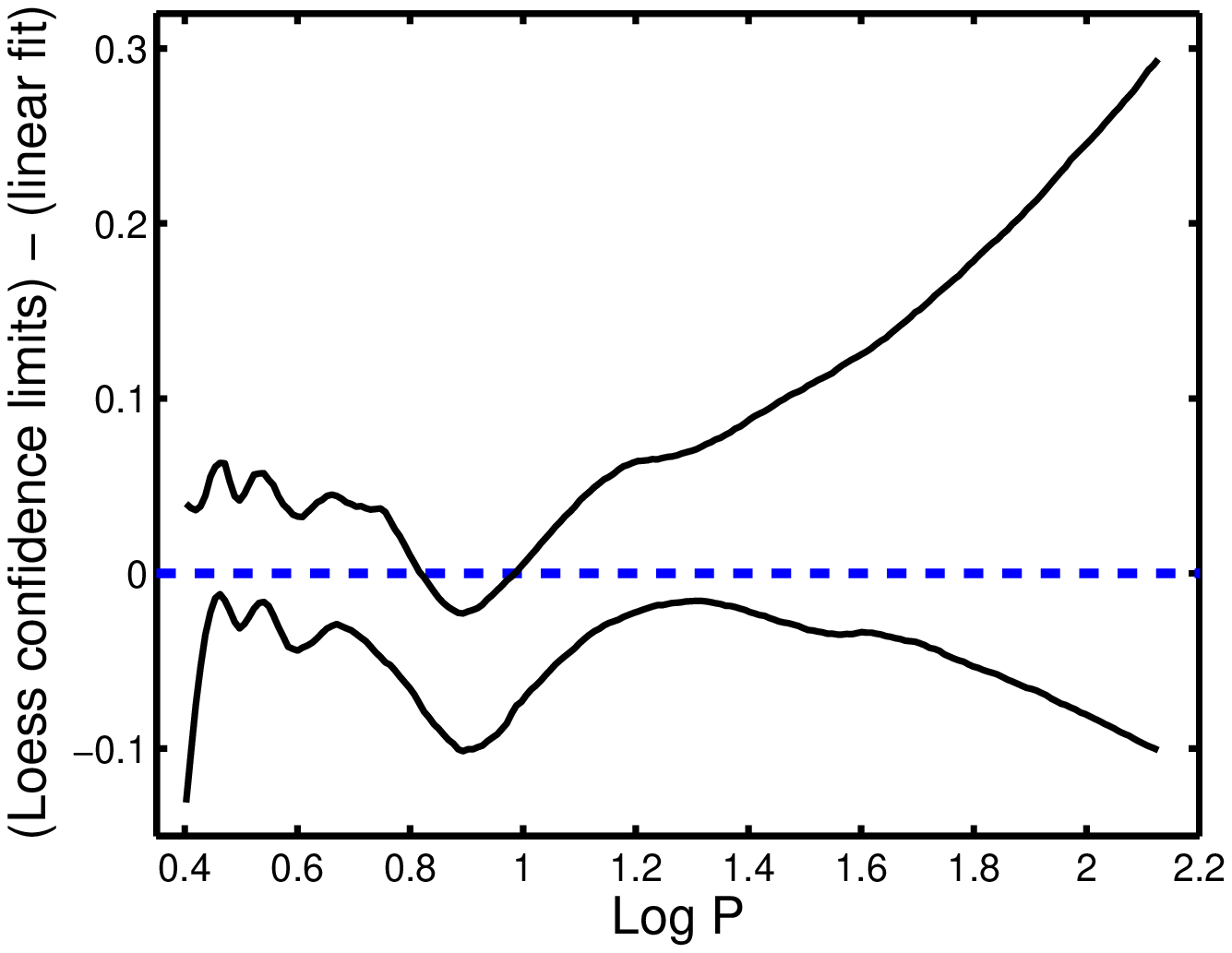}
\caption{The positions (with respect to the linear regression line) 
of the upper and lower 95\% confidence limits
on the loess fit to the OGLE data.}
\end{figure}

\begin{figure}
\epsfysize=8.0cm
\epsffile{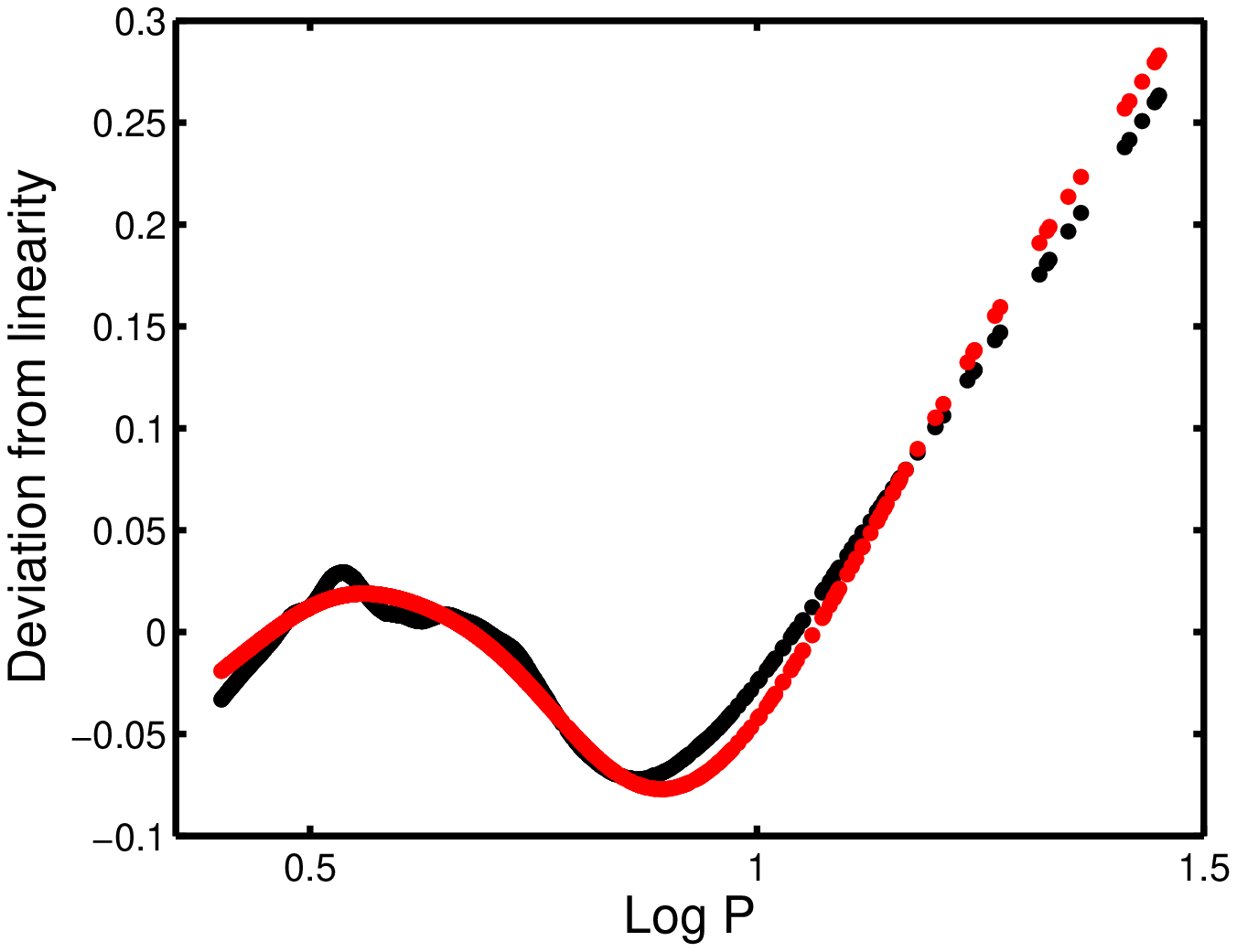}
\caption{Differences between the linear fit and the loess (black, less smooth)
and thin plate regression spline (red, smooth) results for the MACHO data.}
\end{figure}

\begin{figure}
\epsfysize=8.0cm
\epsffile{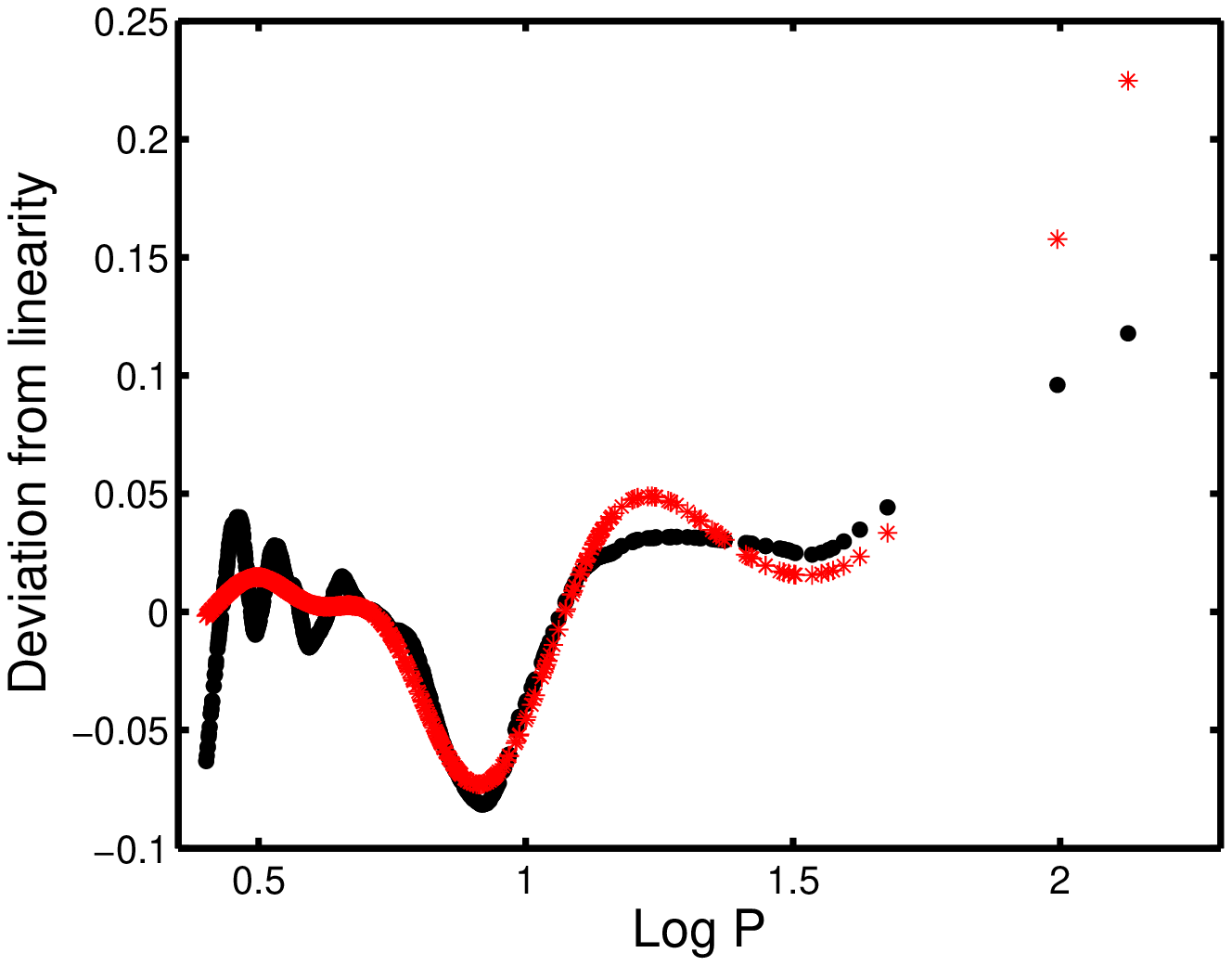}
\caption{Differences between the linear fit and the loess (black, less smooth)
and thin plate regression spline (red, smooth) results for the OGLE data.}
\end{figure}

\begin{figure}
\epsfysize=8.0cm
\epsffile{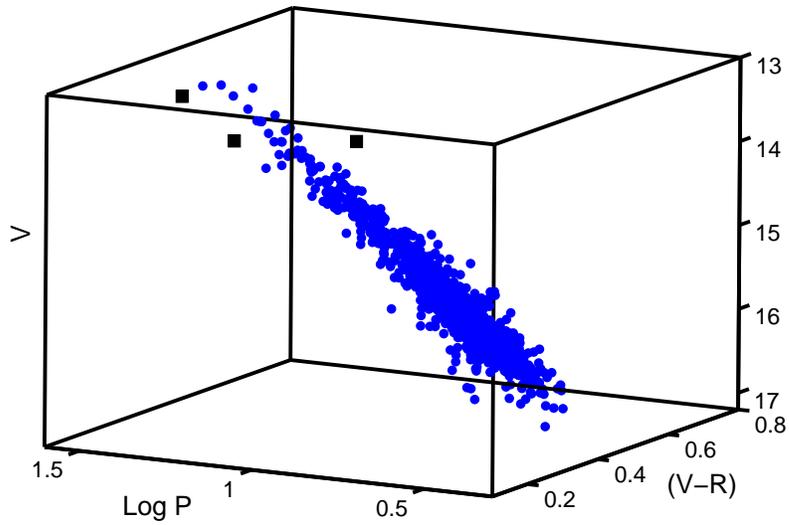}
\caption{The 1216 observations constituting the MACHO dataset. Filled squares mark
the three points selected for deletion on the basis of residual diagnostics.}
\end{figure}

\begin{figure}
\epsfysize=8.0cm
\epsffile{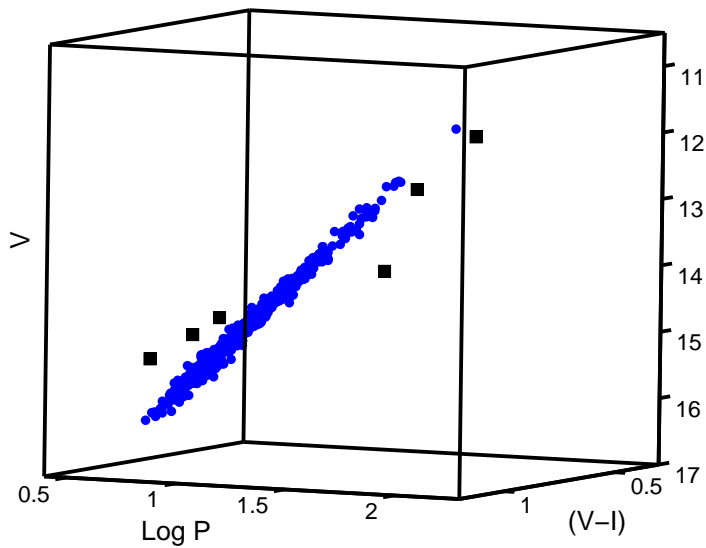}
\caption{The 723 observations constituting the OGLE dataset. Filled squares mark
the six points selected for deletion on the basis of residual diagnostics.}
\end{figure}

\begin{figure}
\epsfysize=8.0cm
\epsffile{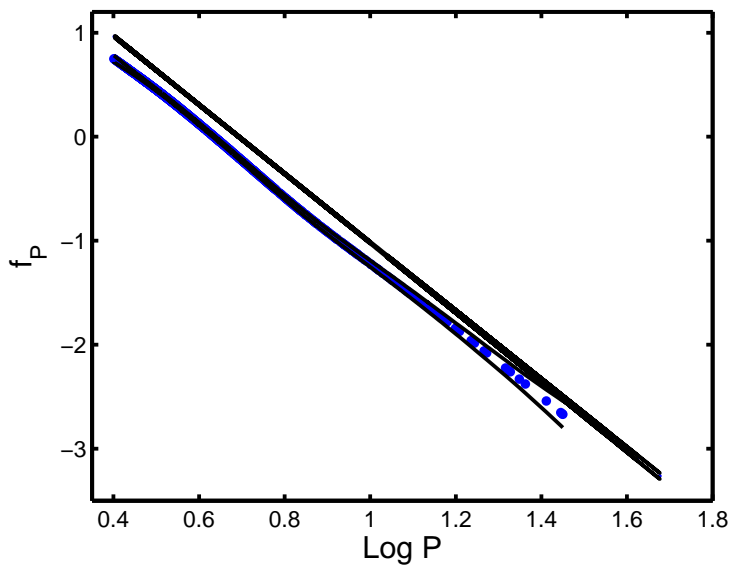}
\caption{The regression functions $f_P$ [see Eqn. (8)] for the OGLE (top)and
MACHO (bottom) data. The $\pm 2$ standard error confidence limits are plotted
as solid lines: these are indistinguishable from the functions except for 
the longer period MACHO data.}
\end{figure}

\begin{figure}
\epsfysize=8.0cm
\epsffile{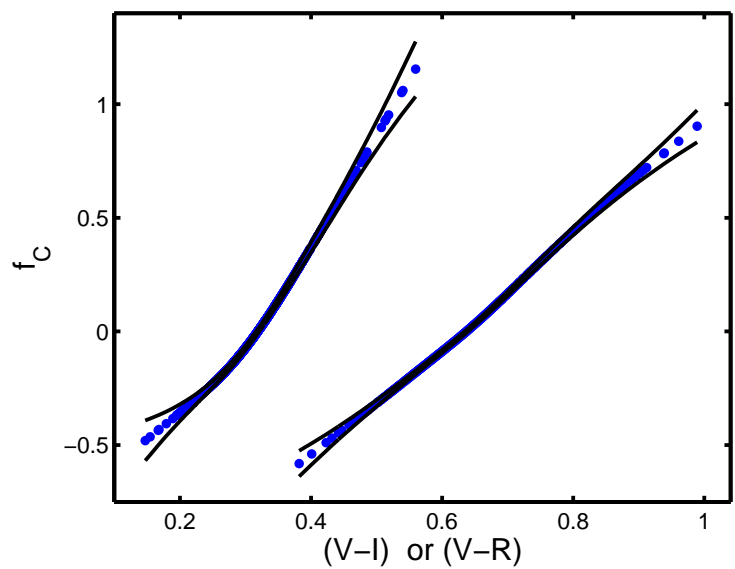}
\caption{The regression functions $f_C$ [see Eqn. (8)] for the MACHO (left) and
OGLE (right) data. The $\pm 2$ standard error confidence limits are plotted
as solid lines.}
\end{figure}

\begin{figure}
\epsfysize=8.0cm
\epsffile{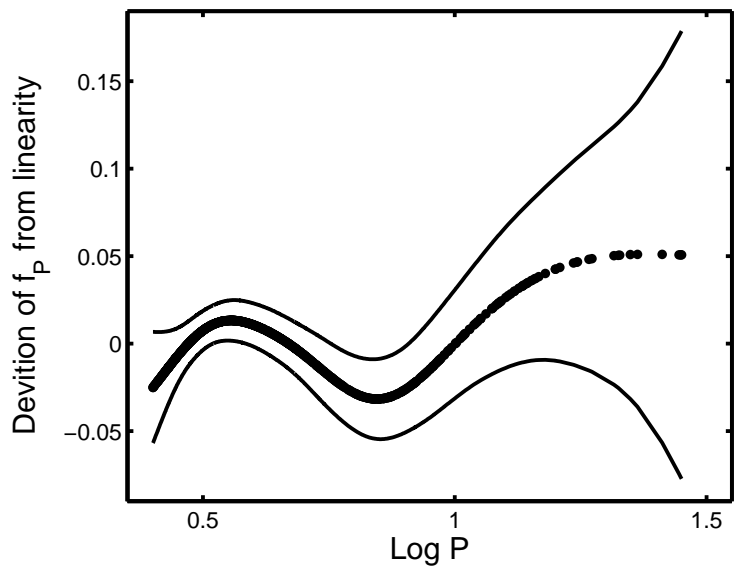}
\caption{The regression functions $f_P$ for the MACHO data (see Fig. 15, bottom plot)
prewhitend by a linear fit, in order to show more clearly the deviations from
linearity. The $\pm 2$ standard error bounds are also plotted.}
\end{figure}

\clearpage

\begin{figure}
\epsfysize=8.0cm
\epsffile{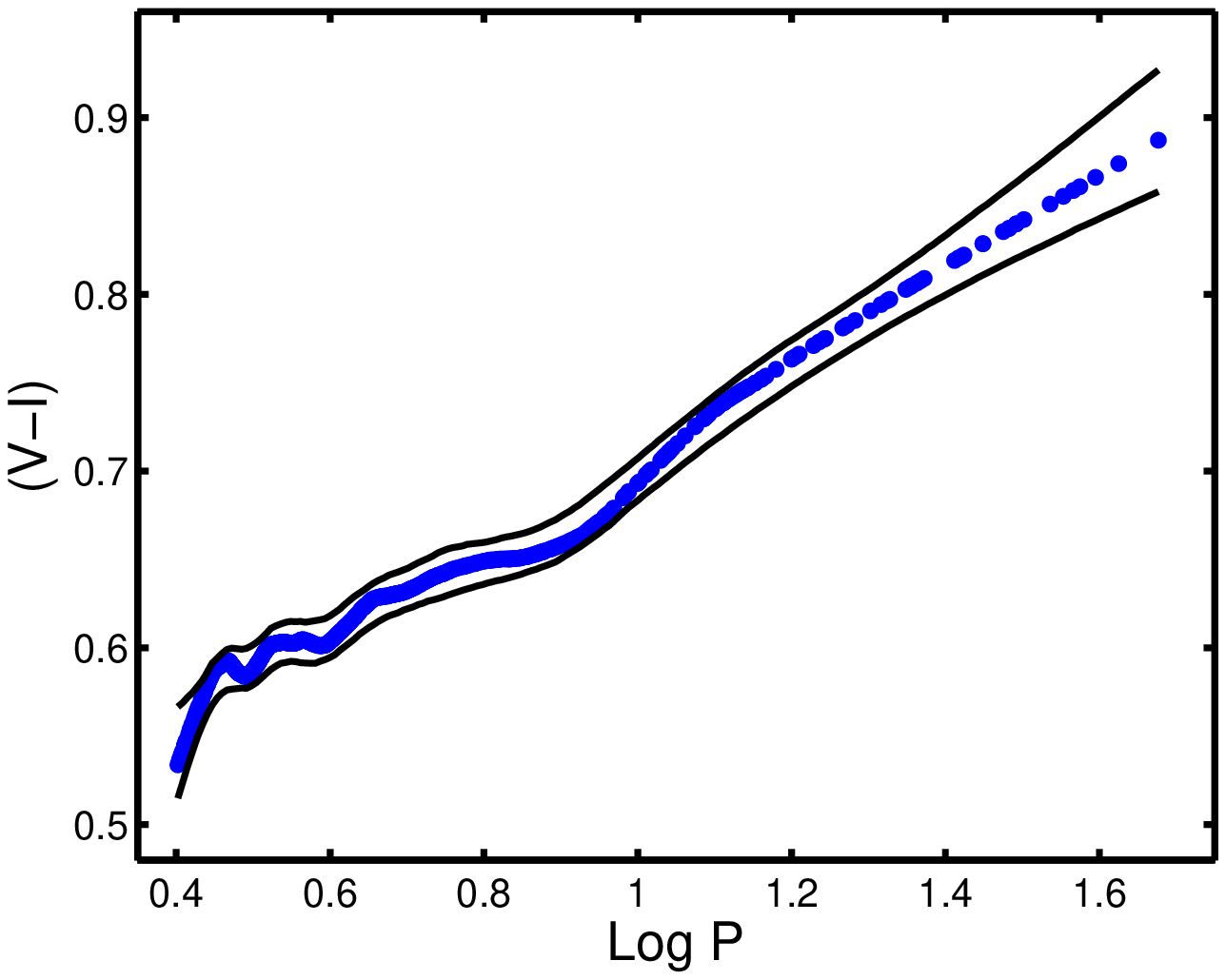}
\caption{A loess regression function fitted to the $\log P$--$(V-I)$ data from
the OGLE observations. The solid lines are the 95\% confidence envelopes, obtained
by bootstrapping.}
\end{figure}

\begin{figure}
\epsfysize=8.0cm
\epsffile{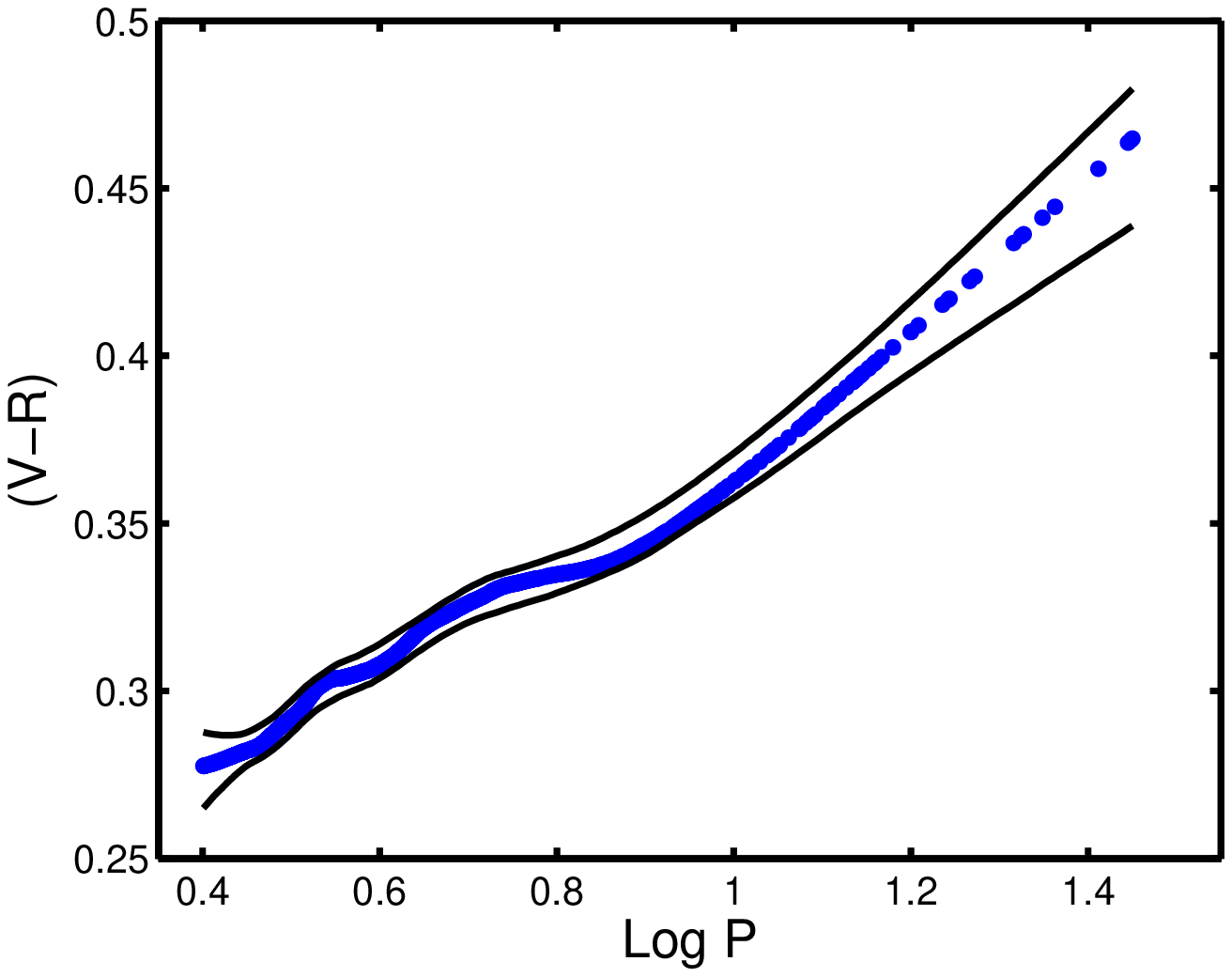}
\caption{A loess regression function fitted to the $\log P$--$(V-R)$ data from
the MACHO observations. The solid lines are the 95\% confidence envelopes, obtained
by bootstrapping.}
\end{figure}

\begin{figure}
\epsfysize=13.0cm
\epsffile{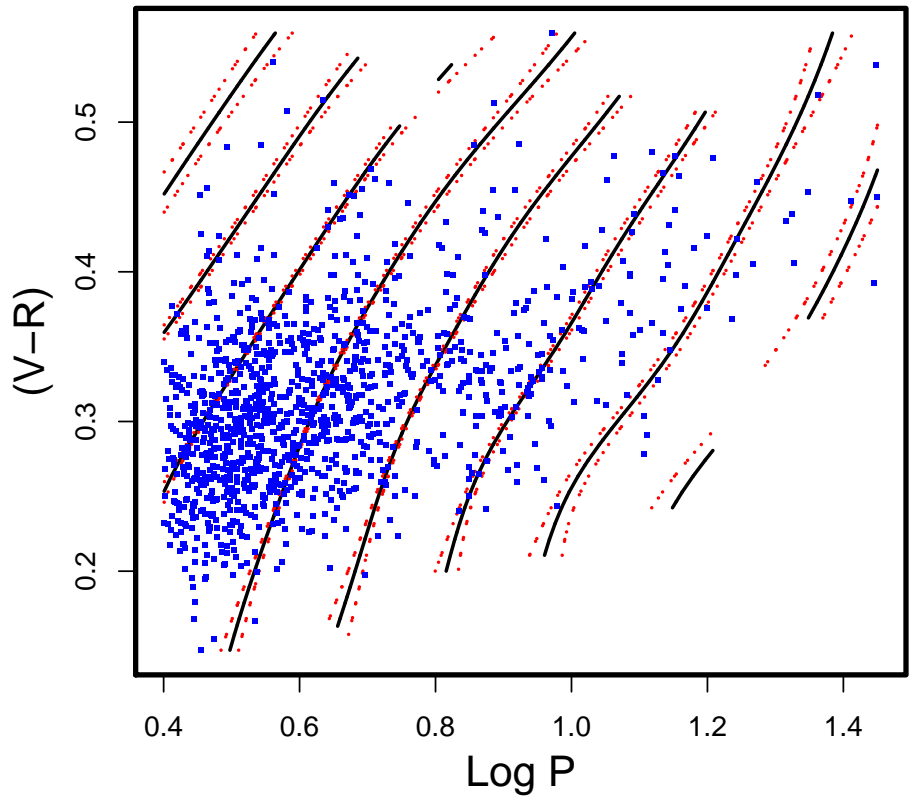}
\caption{A contour plot of the function $f_{PC}$ in (11) fitted to the MACHO data.
The contour values decrease from +1.5 at the top left, in steps of 0.5, to -2 at
the extreme right. The $\pm 1$ standard error bounds for each contour line are also
shown.} 
\end{figure}

\end{document}